\documentclass[conference]{IEEEtran}
\IEEEoverridecommandlockouts
\usepackage{cite}
\usepackage{amsmath,amssymb,amsfonts}
\usepackage{algorithmic}
\usepackage{graphicx}
\usepackage{textcomp}
\usepackage{xcolor}
\def\BibTeX{{\rm B\kern-.05em{\sc i\kern-.025em b}\kern-.08em
    T\kern-.1667em\lower.7ex\hbox{E}\kern-.125emX}}
\begin{document}

\title{A Cross-Platform Mobile Application for Ambulance CPR during Cardiac Arrests
\thanks{This work has been carried out by Omar Alfakir and Viktor Larsson in the context of their Bachelor Thesis at Halmstad University (Computer Science and Engineering program), with the support of Region Halland.}
}

\author{\IEEEauthorblockN{Omar Alfakir, Viktor Larsson, Fernando Alonso-Fernandez}
\IEEEauthorblockA{\textit{School of Information Technology (ITE)} \\
\textit{Halmstad University}, Sweden \\
omaalf18@student.hh.se, viklar18@student.hh.se, feralo@hh.se}
}

\maketitle

\begin{abstract}

This paper describes the implementation of a cross-platform software application to aid ambulance paramedics during CPR (Cardio-Pulmonary Resuscitation). It must be able to work both on iOS and Android devices, which are the leading platforms in the mobile industry. The goal of the application is to guide paramedics in the different processes and expected medication to be administered during a cardiac arrest, a scenario that is usually stressful and fast-paced, thus prone to errors or distractions. 
The tool must provide timely reminders of the different actions to be performed during a cardiac arrest, and in an appropriate order, based on the results of the previous actions. A timer function will also control the duration of each step of the CPR procedure.
The application is implemented in React Native which, using JavaScript as programming language, allows to deploy applications that can run both in iOS and Android native languages.
Our solution could also serve as a record of events that could be transmitted (even in real-time) to the hospital without demanding explicit verbal communication of the procedures or medications administered to the patient during the ambulance trip. This would provide even higher efficiency in the process, and would allow automatic incorporation of the events to the medical record of the patient as well.

\end{abstract}


\section{Introduction}

The integration of healthcare and information technology (healthcare information technology, or HIT) allows to enhance work quality and efficiency in the healthcare sector, contributing to saving people’s lives. 
%
%Incorporating technical solutions in healthcare procedures 
HIT is a promising direction that is showing high value in the optimization of medical applications.
For example, it helps to reduce human errors, facilitating care coordination, improving practice efficiencies, and contributes to enhanced diagnoses in some complex cases where the manual analysis of many different data sources can be difficult. 
HIT has been applied to several medical areas already, although many others still rely on traditional tools and methods, such as paper notes  \cite{1,8}. 

This work is a collaboration of Halmstad University with ambulance personnel in Region Halland (the regional health authority where the university is located). 
The goal is to develop a digital tool to assist personnel in ambulances when treating cardiac arrests. 
During CPR (Cardio-Pulmonary Resuscitation), there are different steps to be performed one after the other during a defined amount of time. The paramedics need to control the flow of events appropriately, while providing an specific amount of medication as well. 
A small distraction in such stressful episodes could stall or make the paramedics forget in which step they are, or what the next is. 
Stress directly impact the performance of paramedics, scoring lower in performance and being unable to report information of the scenario to the same extent \cite{VINCENT2021223}. As such, lowering stress for paramedics can directly influence patient survival rates \cite{11}.
In this context, a digital tool with appropriate reminders and timers could help to keep track of the procedure, reducing the probability of errors and the stress of the involved personnel. 
The application would guide the paramedics through the different steps of the CPR procedure, including timestamps of the operations and medication given. 
Ultimately, it would also increase cardiac arrest patient’s survival rate due to a more efficient CPR process \cite{2}.

\begin{figure}[htb]
\centering
        \includegraphics[width=0.45\textwidth]{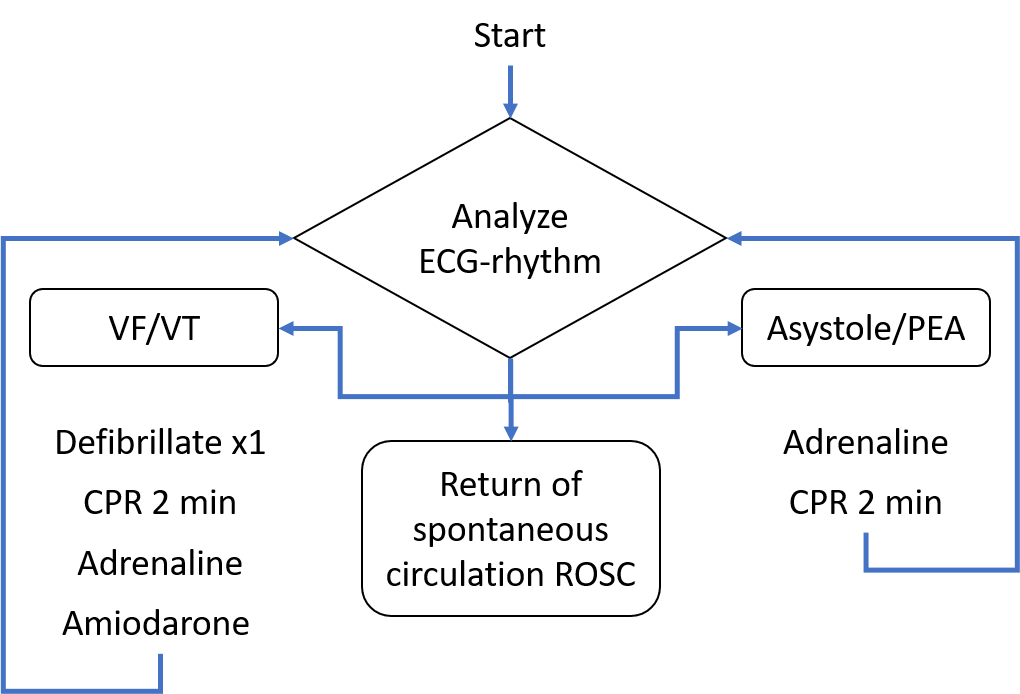}
        \caption{Steps of CPR.}
\label{fig1}
\end{figure}

\begin{figure*}[htb]
\centering
        \includegraphics[width=0.7\textwidth]{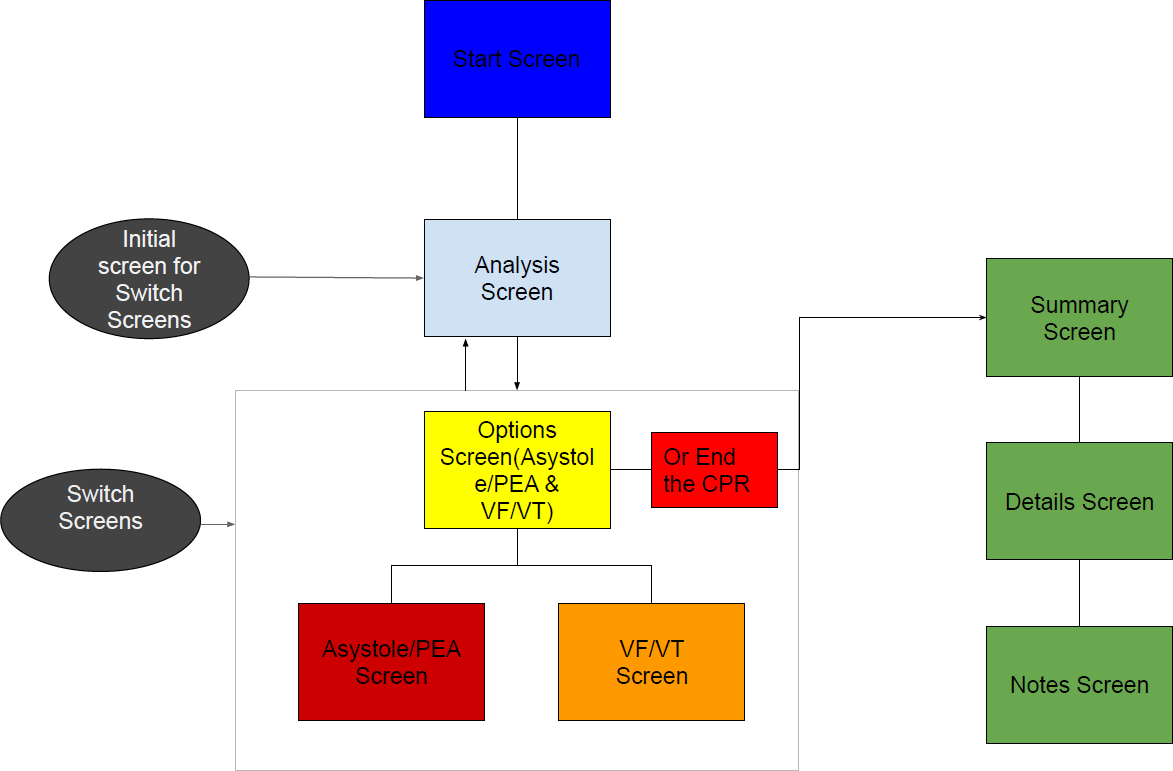}
        \caption{Screens design.}
\label{fig7}
\end{figure*}

\section{Background}

\subsection{CPR (Cardio-Pulmonary Resuscitation) Description}
\label{sect:CPR}

CPR is typically structured around four different states (Figure~\ref{fig1}): Asystole, Pulseless Electrical Activity (PEA), Ventricular Fibrillation (VF) and Ventricular Tachycardia (VT):

\begin{itemize}
    \item Asystole means that the patient is flatlining (no pulse or heart electrical activity).
    \item PEA is a state where the heart’s beat has stopped despite receiving electrical signals from the brain.
    \item VF is when the heart has an abnormal heart rate (the ventricles shake rather than pumping normally).
    \item VT is when the heart beats quickly more than three times in a row and at more than 100 beats per minute.
\end{itemize}

During an Asystole or PEA, the paramedics administer adrenaline (which helps to restart the heart) \cite{12}, and then apply CPR during 2 minutes (consisting of chest compression and rescue breaths). 
Adrenaline is given the first time and then again every four minutes if the heart does not react. 
During VF or VT, the heart state is harmful and causes the brain to panic, causing the heart to remain in this shocking state until it is forced out of it. Therefore, the paramedics use a defibrillator, followed by CPR during 2 minutes. 
After the third defibrillation, they also administer adrenaline and amiodarone (also known as cordarone, which helps to prevent irregular heartbeats), followed by adrenaline every four minutes and amiodarone every other defibrillation. 
%
%VF/VT is the most treatable of cardiac arrest states due to the heart being active \cite{12}.
%
Only one of the four described states can be active at any given time, but a heart attack can start in any of them, and can change to any of them during a cardiac arrest. 
During the process, the Asystole/PEA or VF/VT procedures are administered based on the analysis of the ECG (electrocardiogram). 

\subsection{Related Works}

Existing mobile applications are targeted towards Out-of-Hospital CPR. In other words, towards people who is not knowledgeable about CPR procedures and have no access to medicinal drugs or a defibrillator.
Two popular ones are ''Resuscitate!'' and ''Real Time CPR Guide'' which guide the user through chest compression procedures (with or without rescue breaths). ''Resuscitate'' uses videos to show how to perform CPR, while ''Real-Time CPR Guide'' uses a mixture of images and interactive texts to guide the user. These applications do not offer guides on amounts of medicinal drugs or proper usable documentation, but they are meant to teach a bystander how to perform basic Out-of-Hospital CPR.

The essential aspects that make our application different is the targeted audience and the type of aid that it provides. Our application is mainly meant to help paramedics in proceeding through the different steps of a cardiac arrest as fast and efficiently as possible. 
Its primary purpose is not to show how those steps are performed, but to speed up the decision process and lower the risk of mistakes during CPR. 
In other works, not to be used as a training tool, but to assist personnel who is already highly knowledgeable about CPR.

\section{System Design}

The project is divided into two phases, the documentation
phase and the implementation phase. The documentation
phase aims to find the most suitable setup for this project.
It lays the foundation of how the system will be implemented. 
The implementation phase seeks to describe how the elements 
are integrated to form the intended system.

\subsection{Software Components}

Two platforms currently dominate the mobile market: Android (developed by Google) and iOS (developed by Apple). 
Android is based on Linux, and it has Java and Kotlin as native languages (with Java being the recommended one), supporting C and C++ as well.
iOS, on the other hand, has Objective-C and Swift as native languages. 
This means that, in principle, developers must write applications in each of these native languages if they want to reach the users of both platforms.

To overcome this, several cross-platform mobile frameworks have been developed. They allow to write the application in a single programming language, while being able to generate two versions capable of operating in both platforms.
%
%providing full compatibility with Android and iOS native platforms. 
%
In this work, React Native has been chosen as the programming framework. 
Developed by Facebook in 2015, it is an open-source framework based on JavaScript, aimed at mobile applications development. 
Developers can write the code on JavaScript, and a bridge works as an interpreter to the corresponding native codes. 

Other cross-platform frameworks have been also proposed like Flutter (developed by Google using Dart language), or Xamarin (by Microsoft, using C\#).
React Native has gained attention from companies and developers due to its simplicity and efficiency \cite{9}.
JavaScript is a programming language that is easy to learn, giving a significant advantage in implementing this project in the short time-frame of a bachelor thesis.
It also has a vast community of developers, in comparison for example to Dart (which is a newer language).

\subsection{User Interface (UI)}

The user interface (UI) refers to the visible design and usable functions that the user has access to. UI design is very important in our context, since a poor UI will result in a more difficult interaction, causing unnecessarily delays and a higher probability of errors. 
Accordingly, our priority is to design a clear and simple UI. 
The application needs to be easy to use and understand. 
As such a large part of the focus has been to keep every screen clear of unnecessary information, colors, and buttons, presenting only what is needed. 
Screen buttons are the major part of the interaction between the user and the application, so they need to be big, with the size reflecting its importance or priority of the underlying task. 
For text design, a large and bold text typically reads easily.
Regarding the color scheme, we use bright colors so that when changes are made, they are easily perceived. 
The application’s background is a light grey color, to keep the users focused on the buttons. 
The buttons are in a darker blue color to make them stand out and be of high priority focus when compared to the brighter background and text boxes (Figure~\ref{fig16-23}).

\section{Developed System}

As mentioned earlier, and based upon the paramedic’s desire, our digital solution is in the form of a mobile application that monitors and arranges CPR steps, and gives indications to paramedics based on their previous actions. Certain functions need to be implemented to facilitate paramedics' jobs and reduce human error when documentation during the cardiac arrest event. 
Fundamental functions that the application has been provided with include:

\begin{itemize}
    
    \item An \textbf{alarm function}, which is a countdown timer for two minutes to alert that is time to analyze the ECG of the patient after a CPR procedure. The alarm should produce a sound to alert that the timer value is less than 10 seconds, gestures with warning blinks on the screen from 10 sec to 0, and a vibration when the time is finished.
    
    \item A \textbf{timer function}, which counts the elapsed time from the beginning, providing precise information of how long the patient has been receiving the intervention.
    
    \item A \textbf{documentation function}, which gives the possibility to write notes while doing CPR in the application, instead of using a separate notebook.
    
\end{itemize}

\begin{figure*}[h]
\centering
        \includegraphics[width=0.88\textwidth]{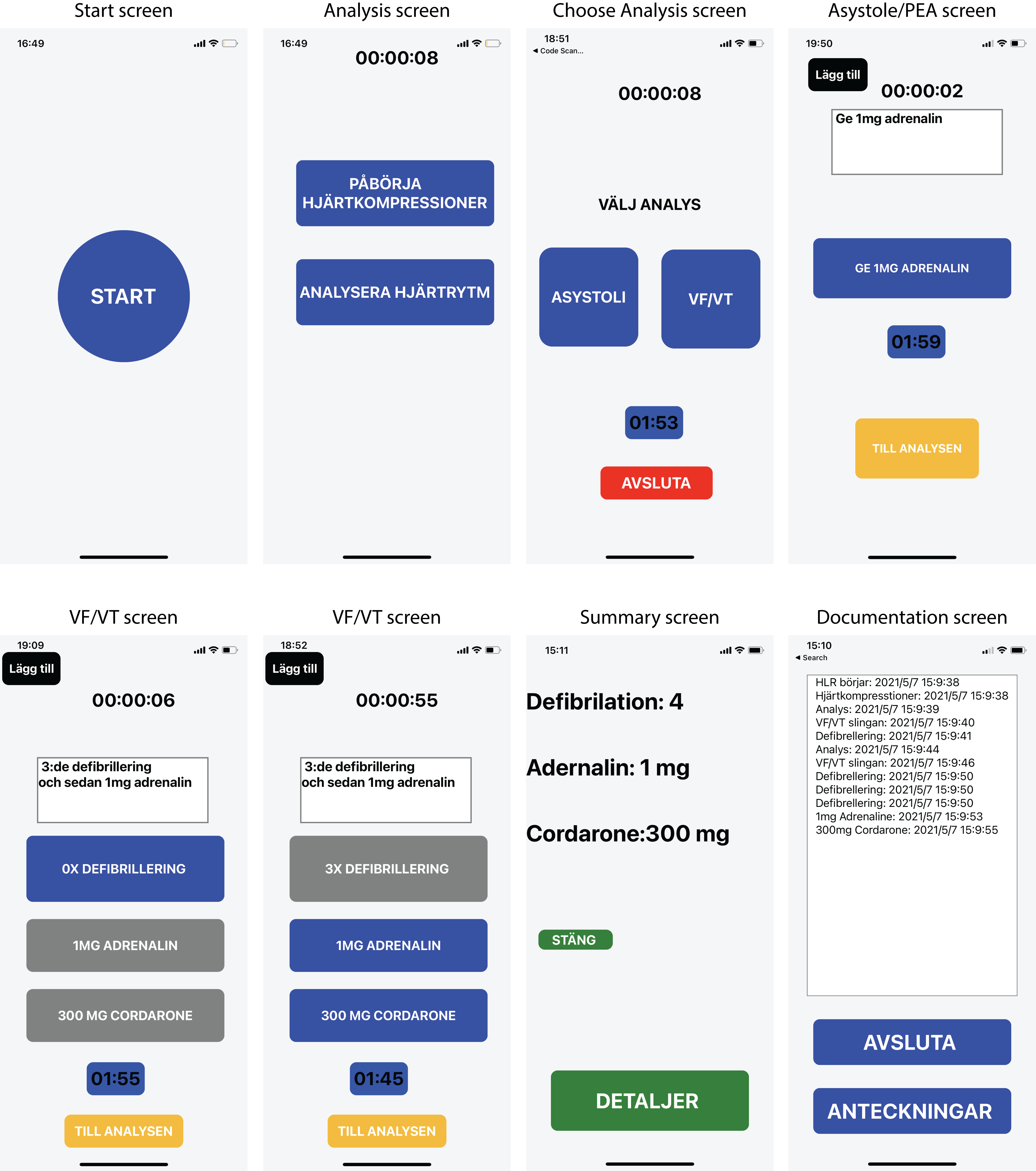}
        \caption{Different screenshots of the application.}
\label{fig16-23}
\end{figure*}

Our application is based on a set of modules or stacks of views (Figure~\ref{fig7}), building up a structure from a base point. 
These stacks are then manipulated depending on the user input, so the user is directed to the appropriate module.
The majority of the user interactions are controlled through buttons, keeping interaction as fast-paced and easy as possible, since the application is to be used in stressful situations. 
It also keeps track of all interactions and choices in the background, mainly through timers and counters.
When new steps are to be taken, appropriate attention is obtained through visual and auditive alarms.
The screen modules of our application, as detailed in Figure~\ref{fig7}, include the following (shown in Figure~\ref{fig16-23}):

\begin{itemize}
    \item \textbf{Start screen} (Figure~\ref{fig16-23}, top left), which contains a start button to initialize the CPR.

    \item \textbf{Analysis screen}, which comes after the user clicks on the start button on the Start screen. This screen contains two buttons, ''Start heart compression'' (during 2 minutes) and ''Analyze heartrythm''. 

    \item \textbf{Choose Analysis screen}, which comes after the analysis screen when the ''Analyze heartrythm'' button is clicked. This screen gives three options: go to the Asystole/PEA screen (top right), go to the VF/VT screen (bottom left), or finish the CPR (''Avsluta'' red button) and go to the Summary blocks. The Asystole/PEA and VF/VT screens contain a yellow button ('''Till Analysen') which conducts to the Analysis screen again.

    \item \textbf{Summary blocks}, which contains a Summary screen that displays the number of defibrillations, and the amount of adrenaline and cordarone given (in mg). This is important because when the patient is transferred from an ambulance to a hospital, the paramedics need to inform the doctors of the amount of medication to avoid accidentally giving the patient more than their body can handle. Then, the Documentation screen lists all the procedures that have been administered to the patient, including a time stamp (for example, 1mg adrenaline at 2021-05-07 15:09). Finally, a Notes screen (not shown) displays the notes that the paramedics have written during the CPR.

\end{itemize}

The application implements the steps described in Section~\ref{sect:CPR}, according to Figure~\ref{fig1}. 
Currently, it stores the data locally. 
It navigates through different screens according to the CPR procedure.
To keep proper track of the order of actions, it uses flags and states stored in a general function container, so they are not reset as the application moves through the different screens, or the current screen is updated. 
%
%Figure~\ref{fig16-19}, first column, shows the analysis screen, while the second column shows the VF/VT screen. 
%
When a button is pressed, it can act in two ways, either it sends the user to a different screen, or the button deactivates itself. 
The time when a button is pressed is also stored, serving as a documentation tool that can be used to present the sequence of steps and the exact time when they have been taken. 
There is also a black button "Lägg till" (add more) in the top left corner of the Asystole/PEA and VF/VT screens which opens a text window over the current screen, where the user can write notes to describe further the steps taken if necessary. 

%An example of the summary screen can be seen in Figure~\ref{fig16-19}, third column, which shows the number of defribillations and the amount of medication given. This is important because when the patient is transferred from an ambulance to a hospital, the paramedics need to inform the doctors of the amount of medication to avoid accidentally giving the patient more than their body can handle.
%
%Finally, Figure~\ref{fig16-19}, fourth column provides a detailed description of the steps taken during the CPR, the amount of medication given, as well as the date of when the step was executed down to seconds.
%

\section{Discussion}

HIT (healthcare information technology) is the integration of information technology systems into healthcare. It involves, among others, software applications that manage medical data, analyzing, storing and retrieving them whenever necessary for the intended purposes. The HIT ecosystem also includes different hardware elements (from medical devices to computers or mobile devices) that interact with software solutions at the appropriate times \cite{1,8}. 
Although significant progress has been made, applying HIT can be very different from one department to another depending on the specific medical procedures and constraints. 

Here, we have developed an application that can aid ambulance paramedics with CPR (Cardio-Pulmonary Resuscitation) during a cardiac arrest. 
The application go through the different actuations to be done during a CPR, including alerts and alarms when it is time to perform the next action. 
In addition, it documents with timestamps all operations that have been conducted.
It is implemented in the form of a mobile application that can work both in Android ad iOS devices, which are the two platforms that currently dominate the mobile market. 
For this purpose, we have employed React Native, a cross-platform programming framework based on JavaScript that allows to transfer the developed code to the corresponding native codes of Android and iOS.
We expect that this application contributes towards minimizing the impact of distractions and reducing the stress of paramedics during such emergency interventions \cite{11}, ultimately increasing patient’s survival rate \cite{2}.

The application include all the basic functionalities, and it operates storing the data locally.
After the essential functions of the application are thoroughly tested, we are working to include other advanced functions and optimizations.
For example, since personal and medical information has to be handled, appropriate data protection measures are necessary. We plan to implement a limited-time function which saves personal data for a limited time (while it has to be handled by the personnel only), after which any personal information is removed, keeping only general information that is relevant for improving the application and the procedures. 
Since data has to be transferred along with patients, we will also add include mechanisms in the application to transfer it securely from the ambulance to the hospital. 
%
%The current user interface (UI) is the standard from React Native, but we are looking forward to implementing 
%
We are also looking into machine learning methods that can exploit usage data of the application in order to find indicators in the CPR process that can help the paramedics to make the intervention more effective. 
On example is finding correlations between the CPR actuation and reported survival rates \cite{TASLIMITEHRANI2016260} or future readmissions \cite{ASHFAQ2019103256}.

\section*{Acknowledgment}

Author F. A.-F. thanks the Swedish Research Council (VR) and the Swedish Innovation Agency (VINNOVA) for funding his research.
 
%\appendix
%\section*{Appendix: React Native Implementation Details}

%\section*{References}

\bibliographystyle{IEEEtran}

%\bibliography{bibliography}

% Generated by IEEEtran.bst, version: 1.12 (2007/01/11)

\end{document}